\documentclass{article}
\usepackage{fullpage}
\usepackage{prettyref,amsmath,amsthm,amssymb,boxedminipage}
\usepackage{algorithm,algorithmic}
\usepackage{pstricks,pst-node,pst-text,pst-coil,pstricks-add,pst-math,pst-xkey}
\usepackage[colorlinks=true]{hyperref}
\usepackage{graphicx}
\newtheorem{theorem}{Theorem}
\newrefformat{theorem}{Theorem~\ref{#1}}
\newrefformat{eq}{Equation~\eqref{#1}}
\newrefformat{chap}{Chapter~\ref{#1}}
\newrefformat{fig}{Figure~\ref{#1}}
\def\xthm[#1][#2][#3]{\newtheorem{#2}[theorem]{#3} \newrefformat{#2}{#3 \ref{#11}}}
\xthm[#][defn][Definition]
\xthm[#][clm][Claim]
\xthm[#][conj][Conjecture]
\xthm[#][prop][Proposition]
\xthm[#][lem][Lemma]
\xthm[#][eg][Example]
\xthm[#][fact][Fact]
\xthm[#][cor][Corollary]
\xthm[#][app][Appendix]
\xthm[#][alg][Algorithm]
\xthm[#][step][Step]
\xthm[#][rem][Remark]

\newrefformat{prpr}{Property \ref{#1}}
\newcommand{\comment}[1]{}

\newcommand{\LR}{\ensuremath{\mathcal{L}}}

\newcommand{\IP}{\ensuremath{\mathcal{P}}}
\newcommand{\viol}{\ensuremath{\mathcal{V}}}
\newcommand{\OPT}{{\mathop{\mathrm{OPT}}}}

\newcommand{\boxy}[2]{\rput(#1){\psframebox*{\phantom{#2}}}\rput(#1){\psframebox{#2}}}

\newcommand{\algo}{\textsc{IteratedSolver}}

\def\algored{\textsc{IteratedReducer}}

\newcommand{\vz}{\ensuremath{\mathbf{0}}}
\newcommand{\vo}{\ensuremath{\mathbf{1}}}

\newcommand{\bs}{\backslash}

\newcommand{\supp}{\mathop{\mathtt{supp}}}
\newcommand{\NP}{\ensuremath{\mathsf{NP}}}
\newcommand{\PP}{\ensuremath{\mathsf{P}}}
\newcommand{\APX}{\ensuremath{\mathsf{APX}}}

\newcommand{\Z}{\mathbb{Z}}
\newcommand{\R}{\mathbf{R}}
\newtheorem*{main1intro}{\prettyref{theorem:main1}}
\newtheorem*{main2intro}{\prettyref{theorem:main2}}
\newtheorem*{main3intro}{\prettyref{theorem:main3}}
\begin{document}

\date{\today}
\title{Approximability of Sparse Integer Programs
}
%\subtitle{Do you have a subtitle?\\ If so, write it here}

%\titlerunning{Short form of title}        % if too long for running head

\author{David Pritchard        \and
        Deeparnab Chakrabarty %etc.
}

%\authorrunning{Short form of author list} % if too long for running head

\comment{\institute{D. Pritchard \at
              Institute of Mathematics, {\'E}cole Polytechnique F\'ed\'erale de Lausanne \\
%              Tel.: +123-45-678910\\
%              Fax: +123-45-678910\\
             \email{david.pritchard@epfl.ch}        %  \\
%             \emph{Present address:} of F. Author  %  if needed
           \and
D. Chakrabarty \at
             Department of Combinatorics and Optimization, University of Waterloo, Waterloo ON N2L 3G1\\ %              Tel.: +123-45-678910\\
%              Fax: +123-45-678910\\
              \email{deeparnab@gmail.com}           %  \\
          }

\date{Received: date / Accepted: date}}
% The correct dates will be entered by the editor

\maketitle

\begin{abstract}
The main focus of this paper is a pair of new approximation algorithms for certain integer programs. First, for covering integer programs  $\{\min cx: Ax \geq b, \vz \leq x \leq d\}$ where $A$ has at most $k$ nonzeroes per row, we give a $k$-approximation algorithm. (We assume $A, b, c, d$ are nonnegative.) For any $k \geq 2$ and $\epsilon>0$, if $\PP \neq \NP$ this ratio cannot be improved to $k-1-\epsilon$, and under the unique games conjecture this ratio cannot be improved to $k-\epsilon$. One key idea is to replace individual constraints by others that have better rounding properties but the same nonnegative integral solutions; another critical ingredient is knapsack-cover inequalities. Second, for packing integer programs $\{\max cx:Ax \leq b, \vz \leq x \leq d\}$ where $A$ has at most $k$ nonzeroes per column, we give a $(2k^2+2)$-approximation algorithm. Our approach builds on the iterated LP relaxation framework. In addition, we obtain improved approximations for the second problem when $k=2$, and for both problems when every $A_{ij}$ is small compared to $b_i$. Finally, we demonstrate a $17/16$-inapproximability for covering integer programs with at most two nonzeroes per column.

%Include keywords, PACS and mathematical subject classification numbers as needed.
%\keywords{Integer programming \and Approximation algorithms \and LP rounding}
% \PACS{PACS code1 \and PACS code2 \and more}
% \subclass{MSC code1 \and MSC code2 \and more}
\end{abstract}

\section{Introduction}
We investigate the following problem: what is the best possible approximation ratio for integer programs where the constraint matrix is sparse? To put this in context we recall a famous result of Lenstra \cite{Lenstra1983}: integer programs with a constant number of variables or a constant number of constraints can be solved in polynomial time. Our investigations analogously ask what is possible if each constraint involves at most $k$ variables, or if each variable appears in at most $k$ constraints.

Rather than consider all integer programs, we consider only packing and covering problems. Such programs have only
positive quantities in their parameters.
One reason for this is that \emph{every} integer program can be rewritten (possibly with additional variables) in such a way that each constraint contains at most 3 variables and each variable appears in at most 3 constraints, if both positive and negative coefficients are allowed. Aside from this, packing programs and covering programs capture a substantial number of combinatorial optimization problems and are interesting in their own right.

A  \emph{covering} (resp.\ \emph{packing}) \emph{integer program}, shorthanded as CIP (resp.\ PIP) henceforth, is
an integer program of the form $\{\min cx: Ax \geq b, \vz \leq x \leq d\}$ (resp.\ $\{\max cx: Ax \leq b, \vz \leq x \leq d\}$) with $A, b, c, d$ nonnegative and rational. Note that CIPs are sometimes called \emph{multiset multicover} when $A$ and $b$ are integral. We call constraints $x \leq d$ \emph{multiplicity constraints} (also known as \emph{capacity constraints}). We allow for entries of $d$ to be infinite, and without loss of generality, all finite entries of $d$ are integral. An integer program with constraint matrix $A$ is \emph{$k$-row-sparse}, or \emph{$k$-RS}, if each row of $A$ has at most $k$ entries; we define \emph{$k$-column-sparse ($k$-CS)} similarly.  As a rule of thumb we ignore the case $k=1$, since such problems trivially admit fully polynomial-time approximation schemes (FPTAS's) or poly-time algorithms.
%{\bf DC: Why FPTAS? Isn't it polytime for all?}
The symbol \vz\ denotes the all-zero vector, and similarly \vo\ denotes the all-ones vector. For covering problems an \emph{$\alpha$-approximation algorithm} returns a feasible solution with objective value at most $\alpha$ times optimal; for packing, the algorithm returns a feasible solution with objective value is at least $1/\alpha$ times optimal. We use $n$ to denote the number of variables and $m$ the number of constraints (i.e.\ the number of columns and rows of $A$, respectively). Throughout the paper, $A$ will be used as a matrix. We let $A_j$ denote the $j$th column of $A$, and let $a_i$ denote the $i$th row of $A$.

\subsection{$k$-Row-Sparse Covering IPs}
The special case of 2-RS CIP where $A, b, c, d$ are 0-1 is the same as Min Vertex Cover, which is \APX-hard. More generally, 0-1 $k$-RS CIP is the same as $k$-Bounded Hypergraph Min Vertex Cover (a.k.a.\ Set Cover with maximum frequency $k$) which is not approximable to $k-1-\epsilon$ for any fixed $\epsilon>0$ unless \PP=\NP~\cite{DGKR05} ($k-\epsilon$ under the unique games conjecture \cite{KR08}). This special case is known to admit a matching positive result: set cover with maximum frequency $k$ can be $k$-approximated by direct rounding of the naive LP~\cite{Ho82} or local ratio/primal-dual methods \cite{BE81}.

The following results are known for other special cases of $k$-RS CIP with multiplicity constraints: Hochbaum \cite{HH86} gave a $k$-approximation in the special case that $A$ is 0-1; Hochbaum et al.\ \cite{HMNT93} and Bar-Yehuda \& Rawitz \cite{BR01} gave pseudopolynomial 2-approximation algorithms for the case that $k=2$ and $d$ is finite. For the special case $d = \vo$,  Carr et al.\ \cite[\S 2.6]{CFLP00} gave a $k$-approximation, and Fujito \& Yabuta \cite{FY04} gave a primal-dual $k$-approximation. Moreover \cite{CFLP00,FY04} claim a $k$-approximation for general $d$, however, the papers do not give a proof and we do not see a straightforward method of extending their techniques to the general $d$ case.
Our first main result, given in \prettyref{sec:krs-cip}, is a simple proof of the same claim.

\begin{theorem}\label{theorem:main1}
There is a polynomial time $k$-approximation algorithm for $k$-RS CIPs with multiplicity constraints.
\end{theorem}
Our approach is to first consider the special case that there are no multiplicity constraints (i.e.\ $d_j = +\infty$ for all $j$); we then extend to the case of finite $d$ via \emph{knapsack-cover inequalities}, using linear programming (LP) techniques from Carr et al.~\cite{CFLP00}. A $(k+1)$-approximation algorithm is relatively easy to obtain using LP rounding; in order to get the tighter ratio $k$, we replace constraints by other ``$\Z_+$-equivalent" constraints (see \prettyref{defn:zeq}) with better rounding properties. The algorithm requires a polynomial-time linear programming subroutine.

Independent simultaneous work of Koufogiannakis \& Young \cite{KY09i,KY09p,KY09d} also gives a full and correct proof of \prettyref{theorem:main1}. Their \comment{primal-iterative} approach works for a broad generalization of $k$-RS CIPs and runs in \comment{low-degree} strongly polynomial time. Our approach has the generic advantage of giving new ideas that can be used in conjunction with other LP-based methods, and the specific advantage of giving integrality gap bounds (see \prettyref{sec:sparsegapbounds}).

\subsection{$k$-Column-Sparse Packing IPs}\label{sec:cspipintro}
Before 2009, no constant-factor approximation was known for $k$-CS PIPs, except in special cases. If every entry of $b$ is $\Omega(\log m)$ then randomized rounding provides a constant-factor approximation. \emph{Demand matching} is the special case of 2-CS PIP where (i) in each column of $A$ all nonzero values in that column are equal to one another and (ii) no two columns have their nonzeroes in the same two rows. Shepherd \& Vetta~\cite{SV07} showed demand matching is \APX-hard but admits a $(\frac{11}{2}-\sqrt{5})$-approximation algorithm when $d = \vo$; their approach also gives a $\frac{7}{2}$-approximation for 2-CS PIP instances satisfying (i). Results of Chekuri et al.\ \cite{CMS07} yield a $11.542k$-approximation algorithm for $k$-CS PIP instances satisfying (i) and such that the maximum entry of $A$ is less than the minimum entry of $b$.

The special case of $k$-CS PIP where $A, b$ are 0-1 is the same as \emph{min-weight $k$-set packing},
\emph{hypergraph matching with edges of size $\leq k$}, and \emph{strong independent sets in hypergraphs with degree at most $k$}.
The best approximation ratio known for this problem is $(k+1)/2+\epsilon$ \cite{Berman00} for general weights,
and $k/2+\epsilon$ when $c = \vo$~\cite{HS89}.
The best lower bound is due to Hazan et al.\ \cite{HSS06}, who showed $\Omega(k/\ln k)$-inapproximability unless
\PP=\NP, even for $c={\bf 1}$. %The natural LP has an integrality gap of at least $(k+1)/2$ even if $c=\vo$: let the ground set be %$\tbinom{[k+1]}{2}$, the sets be $\{\{i, j\} \mid j \in [k+1] \bs \{i\}\}_{i \in [k+1]}.$
%For $k=3$ a 2-approximation in the weighted case is known due to Chan and Lau \cite{CL09}. ONLY 3-dim matching!

Our second main result, given in \prettyref{sec:awesome}, is the following result.
\begin{theorem}\label{theorem:main2}
There is a polynomial time $(2k^2+2)$-approximation algorithm for $k$-CS PIPs with multiplicity constraints.
\end{theorem}
We use the \emph{iterated LP relaxation}~\cite{Singh08} technique to find an integral solution whose objective value is larger than the optimum, but violates some constraints. However the violation can be bounded.
Then we use a colouring argument to decompose the violating solution into $O(k^2)$ feasible solutions giving us the $O(k^2)$-factor algorithm.

The original arXiv eprint and conference version~\cite{P09} of this work gave a $O(k^22^k)$-approximation for $k$-CS PIP using iterated relaxation plus a randomized decomposition approach; that was the first approximation algorithm for this problem with ratio that depends only on $k$. Subsequently in April 2009,  C.~Chekuri, A.~Ene and N.~Korula (personal communication) obtained an $O(k2^k)$ algorithm using randomized rounding, and an $O(k^2)$-approximation in May 2009. The latter method was independently re-derived by the authors, which appears in this version. Finally, Bansal et al.~\cite{BKN09}, in August 2009, gave a simple and elegant $O(k)$-approximation algorithm based on randomized rounding with a careful alteration argument.

\subsection{$k$-Column-Sparse Covering IPs}
Srinivasan \cite{Sri99,Sri06} showed that $k$-CS CIPs admit a $O(\log k)$-approximation.
Kolliopoulos and Young \cite{KY05} extended this result to handle multiplicity constraints. There is a matching hardness result: it is \NP-hard to approximate $k$-Set Cover, which is the special case where $A, b, c$ are 0-1, better than $\ln k - O(\ln \ln k)$ for any $k \geq 3$~\cite{Trev01}. Hence for $k$-CS CIP the best possible approximation ratio is $\Theta(\log k)$. A $(k+\epsilon)$-approximation algorithm can be obtained by separately applying an approximation scheme to the knapsack problem corresponding to each constraint. Although 0-1 2-CS CIP is Edge Cover which lies in \PP, general 2-CS CIP is \NP-hard due to
Hochbaum \cite{Hochbaum04}, who also gave a bicriteria approximation algorithm. Here, we give a stronger inapproximability result.
\begin{theorem}\label{theorem:main3}
For every $\epsilon>0$ it is \NP-hard to approximate 2-CS CIPs of the form $\{\min c\cdot x \mid Ax \geq b, x \textrm{ is } 0\textrm{-}1\}$ and $\{\min c\cdot x \mid Ax \geq b, x \geq 0, x \textrm{ integral}\}$ within ratio $17/16-\epsilon$ even if the nonzeroes of every column of $A$ are equal and $A$ is of the block form $\bigl[ \begin{smallmatrix} A_1 \\ A_2 \end{smallmatrix}\bigr]$ where each $A_i$ is 1-CS.
\end{theorem}
Our proof modifies a construction of \cite{CG08}; we also note a construction of \cite{SV07} can be modified to prove \APX-hardness for the problem.

\subsection{Other Work}
The special case of 2-RS PIP where $A, b, c$ are 0-1 is the same as Max Independent Set, which is not approximable within $n/2^{\log^{3/4+\epsilon} n}$ unless $\NP \subset \mathsf{BPTIME}(2^{\log^{O(1)} n})$
\cite{KP06}. On the other hand, $n$-approximation of any packing problem is easy to accomplish by looking at the best singleton-support solution. A slightly better $n/t$-approximation, for any fixed $t$, can be accomplished by exhaustively guessing the $t$ most profitable variables in the optimal solution, and then solving the resulting $t$-dimensional integer program to optimality via Lenstra's result \cite{Lenstra1983}.

A closely related problem is $k$-Dimensional Knapsack, which are PIPs or CIPs with at most $k$ constraints (in addition to nonnegativity and multiplicity constraints). For fixed $k$, such problems admit a PTAS and pseudo-polynomial time algorithms, but are weakly \NP-hard; see \cite{KPP04} and \cite[Ch. 9]{P09thesis} for detailed references.

\label{sec:semimod}
When $d=\vo$, a natural way to generalize CIP/PIPs is to allow the objective function to be submodular (rather than linear). For minimizing a submodular objective subject to $k$-row sparse covering constraints, the framework of Koufogiannakis \& Young \cite{KY09i,KY09p,KY09d} gives a $k$-approximation; if also $A, b$ are 0-1 (i.e.\ submodular set cover) Iwata and Nagano~\cite{IN09} give a $k$-approximation for all $k$ and Goel et al.~\cite{GKTW09} give a 2-approximation for $k=2$. For maximizing a monotone submodular function subject to $k$-column sparse packing constraints, the algorithm of Bansal et al.~\cite{BKN09} gives a $O(k)$-approximation algorithm.

\subsection{Summary}
We summarize our results and preceding ones in Table~\ref{tab1}; recall also the follow-up $O(k)$ approximation for $k$-CS PIPs~\cite{BKN09}. Note that in all four cases, the strongest known lower bounds are obtained even in the special case that $A, b, c, d$ are 0-1.

\begin{table}[ht]
\begin{center}
\begin{tabular}{||c|cc|ccc||}
\hline \hline
& \multicolumn{2}{c}{$k$-Column-Sparse} & \multicolumn{2}{c}{$k$-Row-Sparse}&\\
& lower bound & upper bound & lower bound & upper bound&\\
\hline
Packing & $\Omega(k / \ln k)$ & $\mathbf{2k^2+2}$ & $n^{1-o(1)}$ & $\epsilon n$& \\
Covering & $\ln k - O(\ln \ln k)$ & $O(\ln k)$ & $k-\epsilon$ & $\mathbf{k}$& \\
\hline \hline
\end{tabular} \caption{The landscape of approximability of sparse integer programs. Our main results are in boldface.}\label{tab1}
\end{center}
\end{table}

\section{$k$-Approximation for $k$-Row-Sparse CIPs}\label{sec:krs-cip}
By scaling rows suitably and clipping coefficients that are too high (i.e.\ setting $A_{ij} = \min\{1, A_{ij}\}$), we may make the following assumption without loss of generality.
\begin{defn}
A \emph{$k$-RS CIP} is an integer program $\{\min c\cdot x: Ax \geq \vo, \vz \leq x \leq d, x\in \Z\}$ where $A$ is $k$-RS and all entries of $A$ are at most 1.
\end{defn}

To begin with, we focus on the case $d_j = +\infty$ for all $j$, which we call the \emph{unbounded $k$-RS CIP}, since it  illustrates the essence of our new technique. Let $x$ be a $n$-dimensional vector of variables
and $\alpha$ is a vector of real coefficients. Throughout, we assume coefficients are nonnegative. When we apply $\lfloor \cdot \rfloor$ to vectors we mean the component-wise floor. That is, the $j$th coordinate of $\lfloor \alpha \rfloor$ is
$\lfloor \alpha_j \rfloor$.
\begin{defn}
A constraint $\alpha \cdot x \geq 1$ is \emph{$\rho$-roundable} for some $\rho > 1$ if for all nonnegative real $x,$ $(\alpha \cdot x \geq 1)$ implies $(\alpha \cdot \lfloor \rho x \rfloor \geq 1)$.
\end{defn}
Note that $\rho$-roundability implies $\rho'$-roundability for $\rho'>\rho$. The relevance of this property is explained by the following proposition.
\begin{prop}\label{prop:algable}
If every constraint in an unbounded covering integer program is $\rho$-roundable, then there is a $\rho$-approximation algorithm for the program.
\end{prop}
\begin{proof}
Let $x^*$ be an optimal solution to the program's linear relaxation. Then $c\cdot x^*$ is a lower bound on the cost of any optimal solution. Thus, $\lfloor \rho x^* \rfloor$ is a feasible integral solution with cost at most $\rho$ times optimal.
\end{proof}
We make another simple observation.
\begin{prop}\label{prop:rowsum}
The constraint $\alpha \cdot x \geq 1$ is $(1 + \sum_i \alpha_i)$-roundable.
\end{prop}
\begin{proof}
Let $\rho = (1 + \sum_i \alpha_i).$ Since $\lfloor t \rfloor > t-1$ for any $t$, if $\alpha \cdot x \geq 1$ for a nonnegative $x$, then
$$\alpha \cdot \lfloor \rho x \rfloor \geq \sum_i \alpha_i (\rho x_i - 1) = \rho \sum_i \alpha_i x_i - \sum_i \alpha_i \geq \rho  - (\rho - 1) = 1,$$
as needed.
\end{proof}

Now consider an unbounded $k$-RS CIP. Since each constraint has at most $k$ coefficients, each less than 1, it follows from \prettyref{prop:rowsum} that every constraint in these programs is $(k+1)$-roundable, and so such programs admit a $(k+1)$-approximation algorithm by \prettyref{prop:algable}. It is also clear that we can tighten the approximation ratio to $k$ for programs where the sum of the coefficients in every constraint (row) is at most $k-1$. We now show that rows with sum in $(k-1, k]$ can be replaced by other rows which are $k$-roundable.

\begin{defn}\label{defn:zeq}
Two constraints $\alpha \cdot x \geq 1$ and $\alpha' \cdot x \geq 1$ are \emph{$\Z_+$-equivalent} if for all nonnegative integral $x$, $(\alpha \cdot x \geq 1) \Leftrightarrow (\alpha' \cdot x \geq 1).$
\end{defn}

In other words, replacing a constraint by an $\Z_+$-equivalent constraint doesn't affect the value of the CIP.
%
%$\alpha \cdot x \geq 1$ and $\alpha' \cdot x \geq 1$ are $\Z_+$-equivalent if $\alpha \cdot x \geq 1$ is valid for $\{x : x \geq 0, \alpha' \cdot x \geq 1\}$ \emph{and} $\alpha' \cdot x \geq 1$ is valid for $\{x : x \geq 0, \alpha \cdot x \geq 1\}$.

\begin{prop}\label{prop:neato}
Every constraint $\alpha \cdot x \geq 1$ with at most $k$ nonzero coefficients is $\Z_+$-equivalent to a $k$-roundable constraint.
\end{prop}

Before proving \prettyref{prop:neato}, let us illustrate its use.
\begin{theorem}
There is a polynomial time $k$-approximation algorithm for unbounded $k$-RS CIPs.
\end{theorem}
\begin{proof}
Using \prettyref{prop:neato} we replace each constraint with a $\Z_+$-equivalent $k$-roundable one. The resulting IP has the same set of feasible solutions and the same objective function. Therefore, \prettyref{prop:algable} yields a $k$-approximately optimal solution.
\end{proof}
With the framework set up, we begin the technical part: a lemma, then the proof of \prettyref{prop:neato}.
\begin{lem}\label{lem:tech}
For any positive integers $k$ and $v$, the constraint $\sum_{i=1}^{k-1} x_i + \frac{1}{v}x_k \geq 1$ is $k$-roundable.
\end{lem}
\begin{proof}
Let $\alpha \cdot x \geq 1$ denote the constraint, i.e.\ $\alpha_k = \frac{1}{v}$, $\alpha_i = 1$ for $1 \le i<k$. If $x$ satisfies the constraint, then the maximum of $x_1$, $x_2$, \ldots, $x_{k-1}$ and $\frac{1}{v}x_k$ must be at least $1/k$. If $x_i \geq 1/k$ for some $i \neq k$ then $\lfloor kx_i \rfloor \geq 1$ and so $\alpha\cdot\lfloor kx \rfloor \geq 1$ as needed. Otherwise $x_k$ must be at least $v/k$ and so $\lfloor kx_k \rfloor \geq v$ which implies $\alpha\cdot \lfloor kx \rfloor \geq 1$ as needed.
\end{proof}
\begin{proof}[Proof of \prettyref{prop:neato}]
If the sum of coefficients in the constraint is $k-1$ or less, we are done by \prettyref{prop:rowsum}, hence we assume the sum is strictly greater than $k-1$. Without loss of generality (by renaming) such a constraint is of the form
\begin{equation}\sum_{i=1}^k x_i \alpha_i \geq 1\label{eq:c}\end{equation}
where $\vz < \alpha \leq \vo$, $k-1 < \sum_i \alpha_i \leq k$, and the $\alpha_i$'s are nonincreasing in $i$.

Define the \emph{support} of $x$ to be $\supp(x) := \{i \mid x_i > 0\}$.
We claim that for any two distinct $j,\ell$, $\alpha_j + \alpha_\ell > 1$. Otherwise, the
$\sum_i \alpha_i \le (k-2)+1 = k-1$. Thus, for any feasible integral $x$ with $|\supp(x)|\ge 2$, we have
$\alpha\cdot x \ge 1$.
%
%Now $\alpha_{k-1} + \alpha_k > 1$ since $k-1 < \sum_{i<k-1}\alpha_i + \alpha_{k-1}+\alpha_k \leq \alpha_{k-1} + \alpha_k + (k-2)$. Since the $\alpha_i$ are nonincreasing, $\alpha_i + \alpha_j > 1$ for any $i < k, j \leq k$; more generally, any integral $x \geq 0$ with $|\supp(x)| \geq 2$ must satisfy $\alpha x \geq 1.$
To express the set of \emph{all} feasible integral solutions, let $t$ be the maximum $i$ for which $\alpha_i=1$ (or $t=0$ if no such $i$ exists), let $e_i$ denote the $i$th unit basis vector, and let $v = \lceil 1/\alpha_k \rceil$. Then it is not hard to see that the nonnegative integral solution set to constraint \eqref{eq:c} is the disjoint union
\begin{equation}\begin{split}&\{x \mid x \geq 0, |\supp(x)| \geq 2\} \uplus \{ze_i \mid 1 \leq i \leq t, z \geq 1, z\in {\mathbb Z}\} \\\uplus &\{ze_i \mid t < i < k, z \geq 2, z\in {\mathbb Z}\} \uplus \{ze_k \mid z \geq v, z\in {\mathbb Z}\}.\end{split}\label{eq:fs}\end{equation}
The special case $t=k$ (i.e.\ $\alpha_1 = \alpha_2 = \dotsb = \alpha_k = 1$) is already $k$-roundable by \prettyref{lem:tech}, so assume $t<k$. Consider the constraint
\begin{equation}\sum_{i=1}^{t} x_i + \sum_{i=t+1}^{k-1} \frac{v-1}{v} x_i + \frac{1}{v} x_k \geq 1.\label{eq:c2}\end{equation}
Every integral $x \geq 0$ with $|\supp(x)| \geq 2$ satisfies constraint \eqref{eq:c2}. By also considering the cases $|\supp(x)| \in \{0,1\},$ it is easy to check that constraint \eqref{eq:c2} has precisely \prettyref{eq:fs} as its set of feasible solutions, i.e.\ constraint \eqref{eq:c2} is $\Z_+$-equivalent to $\alpha x \geq 1$.
If $t<k-1$, the sum of the coefficients of constraint \eqref{eq:c2} is $k-1$ or less, so it is $k$-roundable by \prettyref{prop:rowsum}. If $t=k-1$, constraint \eqref{eq:c2} is $k$-roundable by \prettyref{lem:tech}. Thus in either case we have what we wanted.
\end{proof}

\subsection{Multiplicity Constraints}\label{sec:wmc}
We next obtain approximation guarantee $k$ even with multiplicity constraints $x \leq d.$ For this we use \emph{knapsack-cover inequalities}. These inequalities represent residual covering problems when a set of variables is taken at maximum multiplicity. Wolsey \cite{Wo82} studied inequalities like this for 0-1 problems to get a primal-dual approximation algorithm for submodular set cover. The LP we use is similar to what appears in Carr et al.\ \cite{CFLP00} and Kolliopoulos \& Young \cite{KY05}, but we first replace each row with a $k$-roundable one.

Specifically, given a CIP $\{ \min c\cdot x \mid Ax \geq \vo, \vz \leq x \leq d, x\in{\mathbb Z} \}$ with $A, d$ nonnegative, we now define the knapsack cover LP. Note that we allow $d$ to contain some entries equal to $+\infty$; if $d_j = +\infty$ and some $i$ has $A_{ij} = 0$ our convention is that $A_{ij}d_j = 0$. Recall, $a_i$ is the $i$th row of $A$ and $\supp(a_i)$ denotes the set
$\{j: A_{ij} > 0\}$.
For a subset $F$ of $\supp(a_i)$ such that $\sum_{j \in F} A_{ij}d_j < 1$, define $A^{(F)}_{ij} = \min \{A_{ij}, 1-\sum_{j \in F}  A_{ij}d_j\}.$ Following \cite{CFLP00,KY05} we define the \emph{knapsack cover LP} for our problem to be
\begin{multline*}
\textrm{KC-LP} = \Bigl\{ \min c\cdot x: \vz \leq x \leq d; \\ \forall i, \forall F\subset \supp(a_i) \textrm{ s.t. } \sum_{j \in F} A_{ij}d_j < 1: \,\, \sum_{j \not\in F} A^{(F)}_{ij} x_j \geq 1 - \sum_{j \in F} A_{ij}d_j\Bigr\}.\end{multline*}

It is not too hard to check that any integral solution to the CIP satisfies the constraints of KC-LP, and thus the solution to the latter is a lower bound on the value of the CIP.

\begin{main1intro}
There is a polynomial time $k$-approximation algorithm for $k$-RS CIPs.
\end{main1intro}
\begin{proof}
Using \prettyref{prop:neato}, we assume all rows of $A$ are $k$-roundable.
Let $x^*$ be the optimal solution to KC-LP. Define $\widehat{x} = \min\{d, \lfloor kx^* \rfloor\},$ where $\min$ denotes the component-wise minimum. We claim that $\widehat{x}$ is a feasible solution to the CIP, which will complete the proof
since the objective value of $\widehat{x}$ is at most $k$ times the objective value of KC-LP.
In other words, we want to show for each row $i$ that $a_i\cdot \widehat{x} \geq 1$.

Fix any row $i$ and define $F = \{j \in \supp(a_i) \mid x^*_j \geq d_j/k\},$ i.e.\ $F$ is those variables in the constraint that were rounded to their maximum multiplicity. If $F = \varnothing$ then, by the $k$-roundability of $a_i\cdot x \geq 1$, we have that $a_i\cdot\widehat{x} = a_i\cdot \lfloor kx^* \rfloor \geq 1$ as needed. So assume $F \neq \varnothing$.
Note that for $j\in F$, we have $\widehat{x}_j = d_j$ and for $j\notin F$, we have $\widehat{x}_j = \lfloor kx^*_j \rfloor$.

If $\sum_{j \in F} A_{ij}d_j \geq 1$ then the constraint $a_i\cdot \widehat{x} \geq 1$ is satisfied; consider otherwise. Since $\lfloor kx^*_j \rfloor > kx^*_j-1$ for $j \not\in F$, since $x^*$ satisfies the knapsack cover constraint for $i$ and $F$, and since $A^{(F)}_{ij} \leq 1 - \sum_{j \in F} A_{ij}d_j$ for each $j$, we have
\begin{align*}
\sum_{j \not\in F} A^{(F)}_{ij} \widehat{x}_j = \sum_{j \not\in F} A^{(F)}_{ij}\lfloor kx^*_j \rfloor  &\geq k \sum_{j \not\in F} A^{(F)}_{ij} x^*_j - \sum_{j \not\in F} A^{(F)}_{ij} \\&\geq k\Big(1 - \sum_{j \in F} A_{ij}d_j\Big) - \Big|\{j:j \in \supp(a_i) \bs F\}\Big|\Big(1 - \sum_{j \in F} A_{ij}d_j\Big) \\
&=  k\Big(1 - \sum_{j \in F} A_{ij}\widehat{x}_j\Big) - \Big|\{j:j \in \supp(a_i) \bs F\}\Big|\Big(1 - \sum_{j \in F} A_{ij}\widehat{x}_j\Big)\end{align*}

Since $F \neq \varnothing$ and $|\supp(a_i)| \leq k$, this gives $\sum_{j \not\in F} A^{(F)}_{ij} \widehat{x}_j \geq 1 - \sum_{j \in F} A_{ij}\widehat{x}_j.$ Rearranging, and using the fact $(\forall j: A_{ij} \geq A_{ij}^{(F)})$, we deduce $a_i \cdot \widehat{x} \geq 1$, as needed.\\

For fixed $k$, we may solve KC-LP explicitly, since it has polynomially many constraints. For general $k$,
no method is currently known to solve KC-LP in polynomial time. However, one can use the ellipsoid method
to find a solution $x^*$ whose objective is lower than that of KC-LP, and which satisfies the knapsack-cover
constraints corresponding to the set $F = \{j: x^*_j \ge d_j/k\}$. Note that this is all we need for the above analysis.
Details of how the ellipsoid method finds such a solution are given in \cite{CFLP00,KY05}.
%
%we follow the ellipsoid algorithm-based approach of \cite{CFLP00,KY05}: rather than solve KC-LP in polynomial time, we obtain a solution $x^*$ which is optimal for a modified KC-LP having not all knapsack-cover constraints, but at least all those for the the $m$ specific $(i, F)$ pairs (depending on $x^*$) used in our proof; thus we still get a $k$-approximation in polynomial time.
\end{proof}

\subsection{Integrality Gap Bounds}\label{sec:sparsegapbounds}
In discussing integrality gaps for $k$-RS CIP problems, we say that the \emph{naive LP relaxation} of $\{\min c\cdot x \mid Ax \geq b, \vz \leq x \leq d, x\in{\mathbb Z} \}$ is the LP obtained by removing the restriction of integrality.
Earlier, we made the assumption that $A_{ij} \leq b_i$ for all $i, j$; let us call this the \emph{clipping assumption}. The clipping assumption is without loss of generality for the purposes of approximation guarantees, however, it affects the integrality gap of the naive LP for unbounded $k$-RS CIP, as we now illustrate. Without the clipping assumption, the integrality gap of $k$-RS CIP problems can be unbounded as a function of $k$; indeed for any integer $M \geq 1$ the well-known covering problem $\{\min x_1 \mid [M]x_1 \geq 1, 0 \leq x_1\}$ has integrality gap $M$. In instances with the clipping assumption and without multiplicity constraints, the previous methods in this section establish that the integrality gap of the naive LP is at most $k+1$.

Even under the clipping assumption, it is well-known that $k$-RS CIPs with \emph{multiplicity constraints} can have large integrality gaps --- e.g.\ $\{\min x_2 \mid [\begin{smallmatrix}M\\M\end{smallmatrix}]x \geq M+1,~\vz \leq x,~x_1 \leq 1\}$ has integrality gap $M$. For bounded instances, the knapsack-cover inequalities represent a natural generalization of the clipping assumption, namely, we perform a sort of clipping even considering that any subset of the variables are chosen to their maximum extent.

%(Note also that if $d_i = +\infty$ for all $i$, then KC-LP is just the naive LP with the clipping assumption.)

We have seen that KC-LP has integrality gap at most $k+1$ on $k$-RS CIP instances. Our methods also show that if we replace each row with a $k$-roundable one (\prettyref{prop:neato}), then the corresponding KC-LP has integrality gap at most $k$. We are actually unaware of any $k$-RS CIP instance with $k>1$ where the integrality gap of KC-LP (without applying \prettyref{prop:neato}) is greater than $k$; resolving whether such an instance exists would be interesting. Some special cases are understood, e.g. Koufogiannakis and Young \cite{KY09d} give a primal-dual $k$-approximation for $k$-CS PIP in the case $A$ is 0-1, also known as hypergraph $b$-matching.

\section{Column-Sparse Packing Integer Programs}\label{sec:cspip}\label{sec:awesome}
In this section we give an approximation algorithm for $k$-column-sparse packing integer programs with approximation ratio $2k^2+2$. We better results for $k=2$, and for programs with high width (we defer the definition to a later subsection). The results hold even in the presence of multiplicity constraints $x \leq d.$
Broadly speaking, our approach is rooted in the demand matching algorithm of Shepherd \& Vetta~\cite{SV07}; their path-augmenting algorithm can be viewed as a restricted form of \emph{iterated relaxation}, which is the main tool in our new approach. Iterated relaxation yields a solution whose objective value is {\em larger} than the optimum, however, the solution violates some constraints. We then decompose this infeasible solution to a collection of feasible solutions while retaining at least a constant fraction of the objective value.

For a $k$-CS PIP \IP\ let \LR(\IP) denote its linear relaxation $\{\max c\cdot x \mid Ax \leq b, \vz \leq x \leq d\}.$
We use the set $I$ to index the constraints and $J$ to index the variables in our program.
We note a simple assumption that is without loss of generality for the purposes of obtaining an approximation algorithm: $A_{ij} \le b_i$ for all $i,j$. To see this, note that if $A_{ij} > b_i$, then every feasible solution has $x_j=0$ and we can simply delete $x_j$ from the instance.

Now we give our iterated rounding method. Let the term \emph{entry} mean a pair $(i, j) \in I \times J$ such that $A_{ij}>0$.
Our iterated rounding algorithm computes a set $S$ of \emph{special} entries; for such a set we let $A_{S \to 0}$ denote the matrix obtained from $A$ by zeroing out the special entries.

\begin{lem}Given a $k$-CS PIP \IP, we can, in polynomial time, find $S$ and nonnegative integral vectors $x^0, x^1$ with $x^0 + x^1 \le d$ and $x^1 \le \vo$ such that
\begin{enumerate}
\item[(a)] $c\cdot (x^0+x^1) \geq \OPT(\LR(\IP))$
\item[(b)] $\forall i \in I$, we have $|\{j : (i, j) \in S\}| \leq k$
\item[(c)] $Ax^0 + A_{S \to 0}x^1 \leq b$.
\end{enumerate}\label{lem:addvio}
\end{lem}
In particular, since $x^1$ is 0-1, $(x^0+x^1)$ is a solution such that for each row $i$, we have  $a_i\cdot (x^0 + x^1) \le b_i + k\max_j A_{ij}$. We now give the proof of the above lemma.
\begin{proof}[Proof of \prettyref{lem:addvio}]
First, we give a sketch.
Recall that $A_j$ denote the $j$th column of $A$ and $a_i$ denotes the $i$th row of $A$.
 Let $\supp(A_j) := \{i \in I \mid A_{ij} > 0\}$, which has size at most $k$, and similarly $\supp(a_i) := \{j \in J \mid A_{ij} > 0\}$. Let $x^*$ be an extreme optimal solution to $\LR(\IP)$. The crux of our approach
is as follows: if $x^*$ has integral values we have made progress. If not, $x^*$ is a \emph{basic feasible solution} so there is a set of $\supp(x^*)=|J|$ linearly independent tight constraints for $x^*$, so the total number of constraints $|I|$ satisfies $|I| \ge |J|$. By double-counting there is some $i\in I$ with $|\supp(a_i)| \le k$, which is what permits iterated relaxation: we discard the constraint for $i$ and go back to the start.

\prettyref{fig:kcspip} contains pseudocode for our iterated rounding algorithm, \algo.

\begin{figure}[h]\begin{center}\begin{boxedminipage}{15cm}
\algo$(A, b, c, d)$
\begin{algorithmic}[1]
\STATE Let $x^*$ be an extreme optimum of $\{\max cx \mid x \in \R^{J}; \vz \le x \le d; Ax \leq b\}$
\STATE Let $x^0 = \lfloor x^* \rfloor, x^1 = \vz, J' = \{j \in J \mid x^*_j \not\in \Z\}, I' = I$, $S=\varnothing$. \label{step:inite}
\LOOP
\STATE Let $x^*$ be an extreme optimum of $\{\max cx \mid x \in [0,1]^{J'}; Ax^0 + A_{S \to 0}(x+x^1) \leq b\}$
\STATE For each $j \in J'$ with $x^*_j=0$, delete $j$ from $J'$
\STATE For each $j \in J'$ with $x^*_j=1$, set $x^1_j = 1$ and delete $j$ from $J'$ \label{step:inito}
\STATE If $J' = \varnothing$, terminate and return $S, x^0, x^1$
%\If{in $(V', E')$ every vertex has degree exactly $k$}
%\STATE Pick any $e \in E'$, add $e$ to $M$, and delete $e$ from $E'$
%\Else
\FOR{each $i \in I'$ with $|\supp(a_i) \cap J'| \le k$}\label{step:forlo}
\STATE Mark each entry $\{(i, j) \mid j \in \supp(a_i) \cap J'\}$ special and add it in $S$ and delete $i$ from $I'$
\ENDFOR
%\EndIf
\ENDLOOP
\end{algorithmic}
\end{boxedminipage}
\end{center}
\caption{Algorithm for $k$-CS PIP.}\label{fig:kcspip}
\end{figure}

Now we explain the pseudocode. The $x^0$ term can be thought of as a preprocessing step which effectively reduces the general case to the special case that $d = \vo$. The term $x^1 \in \{0, 1\}^J$ grows over time. The set $J'$ represents all $j$ that could be added to $x^1$ in the future, but have not been added yet. The set $I'$ keeps track of constraints that have not been dropped from the linear program so far.

Since $x^*$ is a basic feasible solution we have $|I'| \ge |J'|$ in \prettyref{step:forlo}. Being $k$-CS, each set $|\supp(A_j) \cap I'|$ for $j \in J'$ has size at most $k$. By double-counting, $\sum_{i \in I'} |\supp(a_i) \cap J'| \le k|J'| \le k|I'|$ and so some $i \in I'$ has $|\supp(a_i) \cap J'| \le k$. Thus $|I'|$ decreases in each iteration, and the algorithm has polynomial running time. (In fact, it is not hard to show that there are at most $O(k \log |I|)$ iterations.)

The algorithm has the property that $c\cdot (x^0+x^1+x^*)$ does not decrease from one iteration to the next, which implies property (a). Properties (b) and (c) can be seen immediately from the definition of the algorithm.
\end{proof}

Now we give the proof of the main result in this section. Here and later we abuse notation and identify vectors in $\{0,1\}^J$ with subsets of $J$, with $1$ representing containment. That is, if we have two $0,1$ vectors $y$ and $x$
we let $y\subset x$ denote the fact that $y_i = 1$ implies $x_i=1$.
\begin{main2intro}
There is a polynomial time $(2k^2+2)$-approximation algorithm for $k$-CS PIPs with multiplicity constraints.
\end{main2intro}
\begin{proof}
We use Lemma \ref{lem:addvio} to obtain $x^0$ and $x^1$.
The main idea in the proof is to partition the set $x^1$ into $2k^2+1$ sets which are all feasible (i.e., we get $x^1 = \sum_{j=1}^{2k^2+1} y^j$ for 0-1 vectors $y^j$ each with $Ay^j \le b$). If we can establish the existence of such a partition, then we are done as follows: the total profit of the $2k^2+2$ feasible solutions $x^0, y^1, \dotsc, y^{2k^2+1}$ is $c\cdot (x^0 + x^1) \ge \OPT$, so the most profitable is a $(2k^2+2)$-approximately optimal solution.

Call $j, j' \in x^1$ \emph{in conflict at $i$} if $A_{ij}>0, A_{ij'}>0$ and at least one of $(i, j)$ or $(i, j')$ is special. We claim that if $y \subset x^1$ and no two elements of $y$ are in conflict, then $y$ is feasible; this follows from \prettyref{lem:addvio}(c) together with the fact that $A_{ij} \le b_i$ for all $i,j$. (Explicitly, for each constraint we either just load it with a single special entry, or all non-special entries, both of which are feasible.) In the remainder of the proof, we find a $(2k^2+1)$-colouring of the set $x^1$ such that similarly-coloured items are never in conflict; then the colour classes give the needed sets $y^j$ and we are done.

To find our desired colouring, we create a \emph{conflict digraph} which has node set $x^1$ and an arc (directed edge) from $j$ to $j'$ whenever $j, j'$ are in conflict at $i$ and $(i, j)$ is special. Rewording, there is an arc $(j, j')$ iff some $(i, j) \in S$ and $A_{ij'}>0$. (If $(i, j')$ is also special, this also implies an arc $(j', j)$.) The key observation is that each node $j \in x^1$ has indegree bounded by $k^2$, i.e.\ there are at most $k^2$ choices of $j$ such that $(j, j')$ is an arc: to see this note $\#\{i \mid A_{ij'} > 0\} \le k$, and each $i$ in this set has $\#\{j \mid (i, j) \in S\} \le k$. Now we use the following lemma, which completes the proof.
\begin{lem}\label{lem:foonk}
A digraph with maximum indegree $d$ has a $2d+1$-colouring.
\end{lem}
\begin{proof}
We use induction on the number of nodes in the graph, with the base case being the empty graph. Now suppose the graph is nonempty. The average indegree is at most $d$, and the average indegree equals the average outdegree. Hence some node $n$ has outdegree at most the average, which is $d$. In total, this node has at most $2d$ neighbours. By induction there is a $(2d+1)$-colouring when we delete $n$, then we can extend it to the whole digraph by assigning $n$ any colour not used by its neighbours.
\end{proof}
(We remark that \prettyref{lem:foonk} is tight, e.g.\ arrange $2d+1$ vertices on a circle and include an arc from each vertex to its $d$ clockwise-next neighbours; this directed $K_{2d+1}$ cannot be $2d$-coloured.) This ends the proof of \prettyref{theorem:main2}.
\end{proof}

\subsection{Improvements for $k=2$}
We give some small improvements for the case $k=2$, using some insights due to Shepherd \& Vetta~\cite{SV07}. A 2-CS PIP is \emph{non-simple} if there exist distinct $j, j'$ with $\supp(A_j)=\supp(A_{j'})$ and $|\supp(A_j)|=2$. Otherwise, it is simple. Shepherd and Vetta consider the case when all non-zero entries of a column are equal.
Under this assumption, they get a $3.5$ approximation for $2$-CS PIPs, and a $\frac{11}{2} - \sqrt{5} \approx 3.26$ approximation for such simple $2$-CS PIPs, when $d = \vo$. We extend their theorem as follows.
\begin{theorem}\label{theorem:k2}
There is a deterministic $4$-approximation algorithm for 2-CS PIPs. There is also a randomized $6-\sqrt{5} \approx 3.764$-approximation algorithm for simple 2-CS PIPs with $d = \vo$.
\end{theorem}
\begin{proof}[(Sketch)]
Since we are dealing with a 2-CS PIP, each $\supp(A_j)$ is an edge or a loop on vertex set $I$; we abuse notation and directly associate $j$ with an edge/loop. Consider the initial value of $J'$, i.e.\ after executing \prettyref{step:inite}. Then we claim that the graph $(I, J')$ has at most one cycle per connected component; to see this, note that any connected component with two cycles would have more edges than vertices, which contradicts the linear independence of the tight constraints for the initial basic solution $x^*$.

We modify \algo\ slightly. Immediately after \prettyref{step:inite}, let $M \subset J'$ consist of one edge from each cycle in $(I, J')$, and set $J' := J' \bs M$. Then $M$ is a matching (hence a feasible 0-1 solution) and the new $J'$ is acyclic. Modify the cardinality condition in \prettyref{step:forlo} to $|\supp(a_i) \cap J'| \le 1$ (instead of $\le 2$); since $J'$ is acyclic, it is not hard to show the algorithm will still terminate, and $\forall i \in I$, we have $|\{j : (i, j) \in S\}| \leq 1$.

To get the first result, we use a colouring argument from \cite[Thm.\ 4.1]{SV07} which shows that $x^1$ can be decomposed into two feasible solutions $x^1 = y^1 + y^2$. We find that the most profitable of $x^0, M, y^1, y^2$ is a 4-approximately optimal solution.

For the second result, we instead apply a probabilistic technique from \cite[\S 4.3]{SV07}. They define a distribution over subsets of the forest $x^1$; let $z$ be the random variable indicating the subset. Let $p = \frac{1}{20}(5 + \sqrt{5})$. Say that an edge $ii'$ is \emph{compatible} with $z$ if $z$ neither contains an edge with a special endpoint at $i$, nor at $i'$. The distribution has the properties that $z$ is always feasible for the PIP, $\Pr[j \in z] = p$ for all $j \in x^1$, and $\Pr[\supp(A_j) \textrm{ compatible with } z] \geq p$ for all $j \in x^0$. (Simplicity implies that $x^0$ and $x^1$ have no edge in common, except possibly loops, which is needed here.) Finally, let $w$ denote the subset of $x^0$ compatible with $z$. Then $z + w$ is a feasible solution, and $\mathrm{E}[c(z+w)] \geq pc(x^1+x^0)$. Hence the better solution of $z+w$ and $M$ is a $1+1/p = (6-\sqrt{5})$-approximately optimal solution.
\end{proof}

\subsection{Improvements For High Width}\label{sec:hwid}
The \emph{width} $W$ of an integer program is $\min_{ij} b_i/A_{ij}$, taking the inner term to be $+ \infty$ when $A_{ij} = 0$. Note that without loss of generality, $W \geq 1$. From now on let us normalize each constraint so that $b_i = 1$; then a program has width $\geq W$ iff every entry of $A$ is at most $1/W$.

In many settings better approximation can be obtained
as $W$ increases. For example in $k$-RS CIPs with $b = \vo$, the sum of each row of $A$
is at most $k/W$, so Propositions \ref{prop:algable} and \ref{prop:rowsum} give a $(1+k/W)$-approximation algorithm. Srinivasan \cite{Sri99,Sri06} gave a $(1+\ln(1+k)/W)$-approximation algorithm for unbounded $k$-CS CIPs. Using \emph{grouping and scaling} techniques introduced by Kolliopoulos and Stein \cite{KS97}, Chekuri et al.\ \cite{CMS07} showed that no-bottleneck demand multicommodity flow in a tree, and certain other problems, admit approximation ratio $1+O(1/\sqrt{W})$.
Multicommodity flow in a tree (without demands) admits approximation ratio $1+O(1/W)$~\cite{KPP08}. Motivated by these results, we will prove the following theorem.

\begin{theorem}\label{theorem:width}
There is a polynomial time $1+\frac{2k}{W-k}$-approximation algorithm to solve $k$-column-sparse PIPs with $W>k.$
\end{theorem}
For $W \ge 2k$, \prettyref{theorem:width} implies a $1+O(k/W)$-approximation. For fixed $k \geq 4$ and large $W$ this is asymptotically tight since $1+o(1/W)$-approximation is \NP-hard, by results from \cite{GVY93,KPP08} on multicommodity flows in trees.
After the initial publication of \prettyref{theorem:width}~\cite{P09}, Bansal et al.~\cite{BKN09} gave an algorithm with ratio $16 \textrm{e} \cdot k^{1/\lfloor W \rfloor}$, where $\textrm{e} = 2.718...$.

\begin{proof}[Proof of \prettyref{theorem:width}]
Run \algo. From \prettyref{lem:addvio} we see that $c\cdot (x^0+x^1) \ge \OPT$ and, using the width bound, \begin{equation}A(x^0+x^1) \leq (1 + k/W)\vo.\label{eq:knee}\end{equation} Define $\viol(x)$ by $\viol(x) := \{i \in I \mid a_i \cdot x > 1\}$, e.g.\ the set of violated constraints in $Ax \leq \vo$.

\def\LPR{\ensuremath{\mathcal R}}

We want to reduce $(x^0+x^1)$ so that no constraints are violated. In order to do this we employ a linear program. Let $\chi(\cdot)$ denote the characteristic vector. Our LP, which takes a parameter $\widehat{x}$, is
$$\LPR(\widehat{x}): \max \{cx \mid \vz \leq x \leq \widehat{x}, Ax \leq {\bf 1} - \frac{k}{W} \chi(\viol(\widehat{x}))\}.$$
%This LP is similar to $\{x \in \P \mid x \leq \widehat{x}\}$ except for the $k/W$ term.
We can utilize this LP in an iterated rounding approach, described by the following pseudocode.

\begin{center}\begin{boxedminipage}{8cm}
\algored\
\begin{algorithmic}[1]
\STATE Let $\widehat{x} := x^0+x^1$
\WHILE{$\viol(\widehat{x}) \neq \varnothing$}
\STATE Let $x^*$ be an extreme optimum of $\LPR(\widehat{x})$
\STATE Let $\widehat{x}$ = $\lceil x^* \rceil$
\ENDWHILE
\end{algorithmic}
\end{boxedminipage}
\end{center}

We claim that this algorithm terminates, and that the value of $c\widehat{x}$ upon termination is at least $$\frac{1-k/W}{1+k/W}c\cdot (x^0+x^1) \geq \frac{1-k/W}{1+k/W} \OPT.$$
Once we show these facts, we are done, since the for the final $\widehat{x}$, $\viol(\widehat{x}) = \varnothing$ implies $\widehat{x}$ is feasible. As an initial remark, note that each coordinate of $\widehat{x}$ is monotonically nonincreasing, and so $\viol(\widehat{x})$ is also monotonically nonincreasing.

Observe that $\LPR$ in the first iteration has $\frac{1-k/W}{1+k/W}(x^0+x^1)$ as a feasible solution, by Equation~\eqref{eq:knee}. Next, note that $x$ which is feasible for $\LPR$ in one iteration is also feasible for \LPR\ in the next iteration since $\viol(\widehat{x})$ is monotonically nonincreasing; hence the value of $c\cdot x^*$ does not decrease between iterations.

To show the algorithm terminates, we will show that $\viol(\widehat{x})$ becomes strictly smaller in each iteration. Note first that if $i \not\in \viol(\widehat{x})$, the constraint $a_i \cdot x \leq 1$ is already implied by the constraint $x \leq \widehat{x}.$ Hence $\LPR(\widehat{x})$ may be viewed as having only $|\viol(\widehat{x})|$ many constraints other than the box constraints $0 \leq x \leq \widehat{x}$. Then $x$, a basic feasible solution to $\LPR(\widehat{x})$, must have at most $|\viol(\widehat{x})|$ non-integral variables. In particular, using the fact that the program is $k$-CS, by double counting, there exists some $i \in \viol(\widehat{x})$ such that $\#\{j \mid x^*_j \not\in \Z, A_{ij}>0\} \le k$. Thus (using the fact that all entries of $A$ are at most $1/W$) we have $a_i\cdot \lceil x^* \rceil < a_i\cdot x^* + k(1/W) \leq 1$: so $i \not\in \viol(\lceil x^* \rceil)$, and $\viol(\widehat{x})$ is strictly smaller in the next iteration, as needed.
\end{proof}

\section{Hardness of Column-Restricted 2-CS CIP}\label{sec:hardness}

\begin{main3intro}
It is \NP-hard to approximate 2-CS CIPs of the form $\{\min cx \mid Ax \geq b, x \textrm{ is } 0\textrm{-}1\}$ and $\{\min cx \mid Ax \geq b, x \geq 0, x \textrm{ integral}\}$ within ratio $17/16-\epsilon$ even if the nonzeroes of every column of $A$ are equal and $A$ is of the block form $\bigl[ \begin{smallmatrix} A_1 \\ A_2 \end{smallmatrix}\bigr]$ where each $A_i$ is 1-CS.
\end{main3intro}
\begin{proof}
Our proof is a modification of a hardness proof from \cite{CG08} for a budgeted allocation problem. We focus on the version where $x$ is 0-1; the other version follows similarly with only minor modifications to the proof. The specific problem described in the statement of the theorem is easily seen equivalent to the following problem, which we call \emph{demand edge cover in bipartite multigraphs}: given a bipartite multigraph $(V, E)$ where each vertex $v$ has a demand $b_v$ and each edge $e$ has a cost $c_e$ and value $d_e$, find a minimum-cost set $E'$ of edges so that for each vertex $v$ its demand is satisfied, meaning that $\sum_{e \in E' \cap \delta(v)} d_e \geq b_v$. Our construction also has the property that $c_e = d_e$ for each edge --- so from now on we denote both  $d_e$.

The proof uses a reduction from Max-3-Lin(2), which is the following optimization problem: given a collection $\{x_i\}_i$ of 0-1 variables and a family of three-variable modulo-2 equalities called \emph{clauses} (for example, $x_1 + x_2 + x_3 \equiv 1 \pmod{2}$), find an assignment of values to the variables which satisfies the maximum number of clauses. H{\aa}stad \cite{Ha01} showed that for any $\epsilon>0$, it is \NP-hard to distinguish between the two cases that (1) a $(1-\epsilon)$ fraction of clauses can be satisfied and (2) at most a $(1/2+\epsilon)$ fraction of clauses can be satisfied.

Given an instance of Max-3-Lin(2) we construct an instance of demand edge cover as follows. For each variable $x_i$ there are three vertices ``$x_i$", ``$x_i = 0$" and ``$x_i = 1$"; these vertices have $b$-value $4\deg(x_i)$ where $\deg(x_i)$ denotes the number of clauses containing $x_i$. For each clause there are four vertices labelled by the four assignments to its variables that do \emph{not} satisfy it; for example for the clause $x_1 + x_2 + x_3 \equiv 1 \pmod{2}$ we would introduce four vertices, one of which would be named ``$x_1 = 0, x_2 = 0, x_3 = 0$." These vertices have $b$-value equal to 3. Each vertex ``$x_i = C$" is connected to ``$x_i$" by an edge with $d$-value $4\deg(x_i)$; each vertex $v$ of the form ``$x_{i_1} = C_1, x_{i_2} = C_2, x_{i_3} = C_3$" is incident to a total of nine edges each with $d$-value 1: three of these edges go to ``$x_{i_j} = C_j$" for each $j = 1, 2, 3$. The construction is illustrated in Figure \ref{fig:construction}.

\begin{figure}[htb]
\begin{center} \leavevmode
\begin{pspicture}(-0.5,-0.5)(11.5,10)
\psset{unit=0.75cm}
\psline[linewidth=3pt](0,6)(2,7)
\psline[linewidth=3pt](0,6)(2,5)
\boxy{0,6}{$x_i$}
\boxy{2,7}{$x_i=0$}
\boxy{2,5}{$x_i=1$}
\psline(6,1)(12,12)
\psline(6.2,1)(12.2,12)
\psline(5.8,1)(11.8,12)
\psline(6,1)(12,0)
\psline(6,1.2)(12,0.2)
\psline(6,0.8)(12,-0.2)
\psline(6,3)(12,8)
\psline(6.2,3)(12.2,8)
\psline(5.8,3)(11.8,8)
\psline(6,3)(12,4)
\psline(6,3.2)(12,4.2)
\psline(6,2.8)(12,3.8)
\psline(6,5)(12,4)
\psline(6,5.2)(12,4.2)
\psline(6,4.8)(12,3.8)
\psline(6,5)(12,12)
\psline(6.2,5)(12.2,12)
\psline(5.8,5)(11.8,12)
\psline(6,7)(12,8)
\psline(6,7.2)(12,8.2)
\psline(6,6.8)(12,7.8)
\psline(6,7)(12,0)
\psline(6.2,7)(12.2,0)
\psline(5.8,7)(11.8,0)
\psline(6,9)(12,12)
\psline(6,9.2)(12,12.2)
\psline(6,8.8)(12,11.8)
\psline(6,9)(12,8)
\psline(6,9.2)(12,8.2)
\psline(6,8.8)(12,7.8)
\psline(6,11)(12,4)
\psline(6.2,11)(12.2,4)
\psline(5.8,11)(11.8,4)
\psline(6,11)(12,0)
\psline(6.2,11)(12.2,0)
\psline(5.8,11)(11.8,0)
\boxy{6,11}{$x_i=0$}
\boxy{6,9}{$x_i=1$}
\boxy{6,7}{$x_j=0$}
\boxy{6,5}{$x_j=1$}
\boxy{6,3}{$x_k=0$}
\boxy{6,1}{$x_k=1$}
\psset{framearc=.5}
\boxy{13,12}{$x_i=1,x_j=1,x_k=1$}
\boxy{13,8}{$x_i=1,x_j=0,x_k=0$}
\boxy{13,4}{$x_i=0,x_j=1,x_k=0$}
\boxy{13,0}{$x_i=0,x_j=0,x_k=1$}
\end{pspicture}
\end{center}
\caption{Left: the gadget constructed for each variable $x_i$. The vertices shown as rectangles
have $b$-value $4\deg(x_i)$; the thick edges have $d$-value and cost $4\deg(x_i)$. Right: the gadget constructed for the clause $x_i + x_j + x_k \equiv 0 \pmod{2}$. The vertices shown as rounded boxes have $b$-value 3; the thin edges each have unit $d$-value and cost.} \label{fig:construction}
\end{figure}
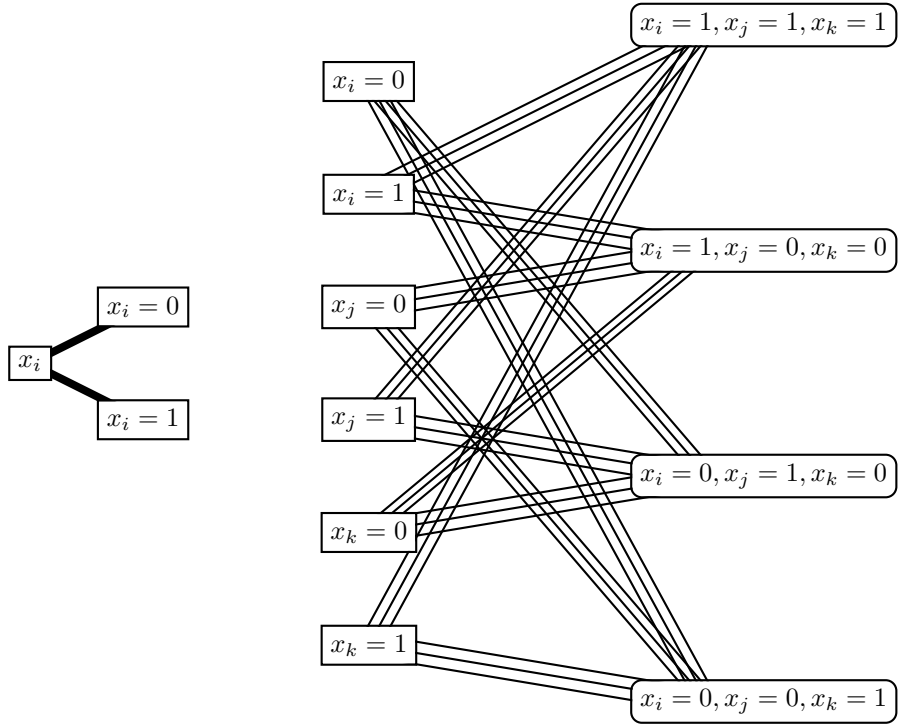

Let $m$ denote the total number of clauses; so $\sum_i \deg(x_i) = 3m$. We claim that the optimal solution to this demand edge cover instance has cost $24m + 3t$ where $t$ is the least possible number of unsatisfied clauses for the underlying Max-3-Lin(2) instance. If we can show this then we are done since H{\aa}stad's result shows we cannot distinguish whether the optimal cost is $\geq 24m + 3m(1/2-\epsilon)$ or $\leq 24m + 3(\epsilon m)$; this gives an inapproximability ratio of $\frac{24+3/2-3\epsilon}{24+3\epsilon} = 17/16 - \epsilon'$ for some $\epsilon' > 0$ such that $\epsilon' \to 0$ as $\epsilon \to 0$, which will complete the proof.

Let $x^*$ denote a solution to the Max-3-Lin(2) instance with $t$ unsatisfied clauses; we show how to obtain a demand edge cover $E'$ of cost $24m+3t$. We include in $E'$ the edge between ``$x_i$" and ``$x_i = x^*_i$" for each $i$; this has total cost $\sum_i 4\deg(x_i) = 12m$. For each satisfied clause $x_i + x_j + x_k \equiv C \pmod{2}$, we include in $E'$ all three edges between ``$x_i = 1-x^*_i$" and ``$x_i = 1-x^*_i, x_j = x^*_j, x_k = x^*_k$" and similarly for $j, k$, and one of each of the parallel triples incident to ``$x_i = 1-x^*_i, x_j = 1-x^*_j, x_k = 1-x^*_k$"; this has cost 12 for that clause. For each unsatisfied clause $x_i + x_j + x_k \equiv C \pmod{2}$, we include in $E'$ any three unit-cost edges incident to ``$x_i = x^*_i, x_j = x^*_j, x_k = x^*_k$," as well as twelve more unit-cost edges: namely in the six nodes consisting of ``$x_i = 1-x^*_i$," ``$x_i = 1-x^*_i, x_j = 1-x^*_j, x_k = x^*_k$" and their images under swapping $i$ with $j$ and $k$, the induced subgraph is a 6-cycle of parallel triples, and we take two edges out of each triple. Thus the chosen edges have total cost 15 for that clause. It is not hard to see that this solution is feasible --- e.g.\ vertices of the form ``$x_i = 1-x^*_i$" are covered by 4 edges for each clause containing them. The total cost is $c(E') = 12m + 12(m-t) + 15t = 24m+3t$.

To finish the proof we show the following.
\begin{clm}
Given a feasible demand edge cover $E'$, we can find a solution $x^*$ such that $t$, the number of unsatisfied clauses for $x^*$, satisfies $24m+3t \leq c(E')$.
\end{clm}
\begin{proof}
First we claim it is without loss of generality that for each $i$, $E'$ contains exactly one of the edges incident to ``$x_i$". Clearly at least one of these two edges lies in $E'$; if both do, then remove one (say, the edge between ``$x_i$" and ``$x_i = 0$") and add to $E'$ any subset of the other $6\deg(x_i)$ edges incident to ``$x_i = 0$" so that the total number of edges incident on ``$x_i = 0$" in $E'$ becomes at least $4\deg(x_i)$. The removed edge has $d$-value $4\deg(x_i)$ and all other incident edges have $d$-value 1, so clearly the solution is still feasible and the cost has not increased.

Define $x^*$ so that for each $i$, $E'$ contains the edge between ``$x_i$" and ``$x_i=x^*_i$." Let $E''$ denote the edges of $E'$ incident on clause vertices (i.e.\ the edges of $E'$ with unit $d$-value). For $F \subset E''$ their \emph{left-contribution}, denoted $\ell(F)$, is the number of them incident on vertices of the form ``$x_i=1-x^*_i$." Note that $\ell(F) \leq |F|$ for any $F$. Furthermore for each unsatisfied clause, all edges incident on its vertex ``$x_i = x^*_i, x_j = x^*_j, x_k = x^*_k$" have zero left-contribution, but $E'$ contains at least three of these edges. Thus the edges of $E''$ incident on that clause's vertices have $\ell(F) \leq |F|-3$. Finally, consider $\ell(E'')$. Each edge of $E''$ is in the gadget for a particular clause, and it follows that $\ell(E'') \leq |E''| - 3t$ where $t$ is the number of unsatisfied clauses for $x^*$. However, $E''$ needs to have $4\deg(x_i)$ edges incident on each ``$x_i=1-x^*_i$" so $\ell(E'') \geq \sum_i 4\deg(x_i) = 12m$. Thus $|E''| \geq 12m + 3t$ and considering the edges incident on the vertices $``x_i"$ we see that $c(E') \geq 24m+3t$.
\end{proof}
This completes the proof of the reduction.
\end{proof}

\section{Open Problems}
It is natural to conjecture that $k$-CS CIP with a submodular objective admits an approximation ratio depending only on $k$, perhaps $O(\ln k)$ matching the best ratio known for linear objectives.

\comment{It is
\item The approximability of $k$-CS CIPs is known so far only to be between $\Omega(k / \ln k)$ and $O(k)$; improving this would be significant, since even for the 0-1 case (hypergraph matching) no $o(k)$ approximation is known.
Is there any general reason why the special 0-1 cases are essentially as hard as the general cases?}

Although 2-RS IPs are very hard to optimize (at least as hard as Max Independent Set), the problem of finding a \emph{feasible} solution to a 2-RS IP is still interesting. Hochbaum et al.\ \cite{HMNT93} gave a pseudopolynomial-time 2-SAT-based feasibility algorithm for 2-RS IPs with finite upper and lower bounds on variables. They asked if there is a pseudopolynomial-time feasibility algorithm when the bounds are replaced by just the requirement of nonnegativity, which is still open as far as we know. It is strongly \NP-hard to determine if IPs of the form $\{x \geq 0 \mid Ax = b\}$ are feasible when $A$ is 2-CS \cite{Hochbaum04}, e.g.\ by a reduction from 3-Partition; but
for IPs where each variable appears at most twice \emph{including} in upper/lower bounds, it appears all that is known is weak \NP-hardness (for example, via the \emph{unbounded knapsack problem} \cite{Lueker75}).

\subsubsection*{Acknowledgement.}

%\begin{acknowledgements}
We would like to thank Glencora Borradaile, Christina Boucher, Stephane Durocher, Jochen K\"onemann and Christos Koufogiannakis for helpful discussions, and the ESA referees for useful feedback.
%\end{acknowledgements}

\end{document}